\documentclass[aps,showpacs,prb,twocolumn]{revtex4}
\usepackage{mathrsfs}
\usepackage{amssymb}
\usepackage{amsmath}
\usepackage{bm}
\usepackage{graphics}
\usepackage{natbib}

\newcommand{\sss}[1]{{\scriptscriptstyle #1}}
\begin{document}
\title{Thermoelectric effect in a parallel double quantum dot structure}
\author{Wei-Jiang Gong$^{1,2}$}
\author{Guo-Zhu Wei$^{1,2}$}

\affiliation{1.
College of Sciences, Northeastern University,
Shenyang 110819, China \\
3. International Centre for Materials Physics, Chinese Academy of
Sciences, Shenyang, 110016, China}
\date{\today}

\begin{abstract}
We discuss the thermoelectric properties assisted by the Fano effect
of a parallel double quantum dot (QD) structure. By adjusting the
couplings between the QDs and leads, we facilitate the nonresonant
and resonant channels for the Fano interference. It is found that at
low temperature, Fano lineshapes appear in the electronic and
thermal conductance spectra, which can also be reversed by an
applied local magnetic flux with its phase factor $\phi=\pi$. And,
the Fano effect contributes decisively to the enhancement of
thermoelectric efficiency. However, at the same temperature, the
thermoelectric effect in the case of $\phi=\pi$ is much more
apparent, compared with the case of zero magnetic flux. By the
concept of Feynman path, we analyze the difference between the
quantum interferences in the cases of $\phi=0$ and $\phi=\pi$. It is
seen that in the absence of magnetic flux the Fano interference
originates from the quantum interference among infinite-order
Feynman paths, but it occurs only between two lowest-order Feynman
paths when $\phi=\pi$. The increase of temperature inevitably
destroys the electron coherent transmission in each paths. So, in
the case of zero magnetic field, the thermoelectric effect
contributed by the Fano interference is easy to weaken by a little
increase of temperature.

\end{abstract}
\pacs{73.23.Hk, 73.50.Lw, 85.80.Fi} \maketitle

\bigskip

\section{Introduction}
The field of thermoelectricity and solid-state thermionics has
recently received renewed attention, due to advances in growth and
fabrication of complex compounds, mesoscopic devices and
nanostructures.\cite{RMP1,ZhangC} The main purpose is to enhance the
efficiency of solid-state thermoelectric devices at a mesoscopic or
nanoscopic scale.\cite{ChenG} As is known, the thermoelectric
efficiency of a solid-state device is described by the figure of
merit $ZT$. $ZT$ is defined as
\begin{equation}
ZT=S^2{\cal G} T/\kappa,
\end{equation}
where $S$, $\cal G$, and $T$ are thermopower, electronic
conductance, and absolute temperature, respectively.
$\kappa=\kappa_{el}+ \kappa_{ph}$ is the thermal conductance, in
which $\kappa_{el}$ is the electron and $\kappa_{ph}$ the phonon
thermal conductance.\cite{Mahan} Besides, the thermal and electronic
conductances for most macroscopic metals at very low or room
temperatures are constrained by the Wiedemann-Franz law,
\begin{equation}
\kappa/{\cal G} T=L_0,
\end{equation}
where $L_0= k_B^2\pi^3/3e^2$ is the Lorenz number with $k_B$ the
Boltzmann constant and $e$ the electron charge.\cite{Grosso} Since
the relationship between these parameters above, it is difficult to
achieve the increment of thermoelectric efficiency in bulk
materials.
\par
The progress of research on the quantum transport through mesoscopic
systems and nanostructures motivates scientists to pay close
attention to the thermoelectric properties of mesoscale or nanosale
structures.\cite{Science1} Recently, there reported a number of
interesting experimental results, and the barrier of $ZT=1$ has been
overcome at high temperatures in such structures. Harman $et$ $al.$
observed $ZT\simeq 1.6$ in a $\rm{PbSeTe}$ based quantum dot
superlattice.\cite{Herman} Venkatasubramanian $et$ $al.$ achieved
$ZT\approx2.4$ in a p-type $\rm{Bi_2Te_3/Sb_2Te_3}$
superlattice.\cite{Siivola} More recently, Kanatzidis and co-workers
found that bulk $\rm{AgPb_{18}SbTe_{20}}$ with internal
nanostructures has $ZT\approx2$ at $T=800 K$.\cite{Hsu} In
nanocrystalline $\rm{BiSbTe}$ bulk alloys $ZT$ reached the value 1.4
at $T =373 K$.\cite{Poudel} And, a 100-fold improvement of $ZT$
compared to the bulk value has been reported lately in $\rm{Si}$
nanowires.\cite{ChenR,Boukai} On the other hand, Lyeo $et$ $al.$
measured the Seebeck coefficient across a junction formed by a
semiconducting substrate and the tip of a scanning transmission
microscope. The consensus is that finding a material with a
thermoelectric figure of merit $ZT\geq4$ would mark a major
technological breakthrough.\cite{Lyeo} The experimental results were
well explained by theoretical workers. It was found that the
enhanced $ZT$ in nanostructures is attributed to the decrease of the
thermal conductance produced by the scattering of phonons off the
structure,\cite{Hicks,Wang} or due to the increase of thermopower
induced by the presence of enhanced densities of states at the Fermi
level.\cite{Zeitler,Oxford,Mahan2}

\par
QD systems are typical nanostructures, since they contain a variety
of interesting quantum transport properties with their potential
applications. Furthermore, based on the work of Mahan and
Sofo,\cite{Mahan2} it can be anticipated that QDs and molecular
junctions are good candidates to explore the thermoelectric
properties of low-dimensional structures, since the $\delta$-like
density of states and small phonon contribution to thermal
conductance in these systems.\cite{Hicks,Mahan2,Triberis}
Consequently, many experimental and theoretical groups have devoted
themselves to the thermoelectric properties of QDs and molecules, as
a result, some interesting phenomena were
reported.\cite{Staring,Bruder,Koch,Kubala,Zianni,Fazio,DongB,Kraw,Kita,Reuter,Yosida,Nguyen,Triberis,Barnas}
First, it was found that in these systems, the characteristics of
level quantization and Coulomb blockade effects indeed lead to novel
thermoelectric features, such as oscillations of the thermopower and
oscillations of the thermal
conductance.\cite{Staring,Bruder,Zianni,XingDY,Sunqf} On the other
hand, the Coulomb interactions in QD devices have a significant
influence on thermoelectric transport coefficients, and lead to
strong violation of the Wiedeman-Franz law.\cite{Kubala,Costi}
Experiments performed on QDs in the Kondo regime reveal a strong
influence of spin correlations on the thermopower.\cite{Reuter}
Also, Murphy $et$ $al.$ demonstrated that violation of the
Wiedeman-Franz law is the main mechanism of an enhanced
thermoelectric efficiency in molecular junctions, which can be
important for possible applications in energy conversion
devices.\cite{Moore} Then, the above results confirm that the
peculiar properties of QDs play important roles in the change of
thermoelectric properties.

\par
As is known, QDs have important characteristics that some QDs can be
coupled to form coupled-QD systems. In comparison with the single-QD
system, coupled QDs present more intricate quantum transport
behaviors, because of the tunable structure parameters and abundant
quantum interference mechanisms. A variety of interesting phenomena
were reported in the past years, such as negative differential
conductance\cite{NDC}, Pauli spin blockade\cite{Pauli},
multi-orbital Kondo effect\cite{Kondo}, Fano effect\cite{Fano},
decoupled molecular states\cite{Decouple}, etc. In view of these
results, one can anticipate the thermoelectric properties of coupled
QDs will be of much interest. Recently, the thermoelectric
properties of the coupled-QD structures have received much
attention.\cite{ChenX,Finch,Wierzbicki,David} Calculations based on
the density-functional formalism indicate that the thermoelectric
efficiency of molecules which exhibit the Fano resonance can be
significantly enhanced.\cite{DFT} And, Yoshida $et$ $al.$ reported
that in the structure of a QD side coupled to a quantum wire, the
interplay between the quantum interference and Kondo effect makes
nontrivial contributions to thermoelectric properties.\cite{Yosida}
Apart from the T-shaped QD system, thermoelectric effects of the
parallel coupled QDs were also studied extensively. Very recently,
Liu $et$ $al.$ have investigated thermoelectric effects in parallel
double QDs attached to two metallic leads, and with a magnetic flux
threading the QD device.\cite{Liu} They arrived at the conclusion
that the figure of merit ZT can be enhanced in the vicinity of the
Fano resonance. Similar conclusion also follows from a recent paper,
where the influence of electron interference in a two-level system
on the maximum thermoelectric power is analyzed.\cite{Linke}
Besides, it was reported that in such structures the interplay
between the Coulomb correlations and interference effects leads to
strong violation of the Wiedemann-Franz law.\cite{Trocha} Based on
the existing results, one can conclude that in coupled QDs, the
quantum interference plays a significant role in modulating the
thermoelectric effect.

\par
Following such a topic, in this work we would like to carry out a
comprehensive analysis about the thermoelectric behaviors assisted
by the Fano effect. To do so, we choose a parallel double-QD
structure, and adjust the QD-lead couplings to realize the
nonresonant and resonant channels for the Fano
interference.\cite{refFano,refIye} Via numerical calculation, we
find that the Fano effect contributes significantly to the
enhancement of the thermoelectric efficiency. Furthermore, in the
cases of zero magnetic flux and finite magnetic flux with
$\phi=\pi$, the Fano interferences play different roles in the
thermoelectric effect. To be precise, at the same temperature, the
thermoelectric effect in the case of $\phi=\pi$ is more robust
compared with the case of zero magnetic flux. By analyzing the
difference between the quantum interferences in these two cases, the
physics pictures are clarified, so that the magnetic-flux dependence
of thermoelectric effects is well explained. Based on the obtained
results, we believe that this work is helpful for understanding
about the thermoelectric effect of double-QD systems.

\begin{figure}
\begin{center}\scalebox{0.6}{\includegraphics{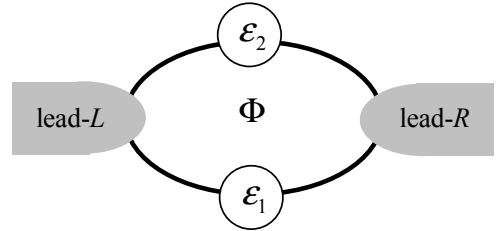}}
\caption{ Schematic of a parallel double-QD structure. $\Phi$
denotes a local magnetic flux through the system. \label{Struct}}
\end{center}
\end{figure}
\section{model\label{theory}}
The parallel double QD structure that we consider is illustrated in
Fig.\ref{Struct}. The Hamiltonian to describe the electronic motion
in this structure reads
\begin{equation}
H=H_{C}+H_{D}+H_{T}.   \label{1}
\end{equation}
The first term is the Hamiltonian for the noninteracting electrons
in the two leads:
\begin{equation}
H_{C}=\underset{\alpha k \sigma}{\sum }\varepsilon _{\alpha
k}c_{\alpha k\sigma}^\dag c_{\alpha k \sigma},\label{2}
\end{equation}
where $c_{\alpha k\sigma}^\dag$ $( c_{\alpha k\sigma})$ is an
operator to create (annihilate) an electron of the continuous state
$|k,\sigma\rangle$ in the lead-$\alpha$ ($\alpha\in L,R$).
$\varepsilon _{k\alpha}$ is the corresponding single-particle
energy. The second term describes electron in the double QDs. It
takes a form as
\begin{equation}
H_{D}=\sum_{j=1,\sigma}^{2}\varepsilon _{j}d_{j\sigma}^\dag
d_{j\sigma},\label{3}
\end{equation}
in which $d^{\dag}_{j\sigma}$ $(d_{j\sigma})$ is the creation
(annihilation) operator of electron in QD-$j$. $\varepsilon_j$
denotes the electron level in the corresponding QD. We assume that
only one level is relevant in each QD. Here in order to present the
leading thermoelectric results affected by the quantum interference,
we ignore the electron interaction in QDs. The last term in the
Hamiltonian describes the electron tunneling between the leads and
QDs, which is given by
\begin{equation}
H_{T} =\underset{\alpha k j\sigma}{\sum }(
V_{j\alpha}d_{j\sigma}^\dag c_{\alpha k \sigma}+{\mathrm {H.c.}}).
\label{4}
\end{equation}
In this equation, $V_{j\alpha}$ denotes the QD-lead coupling
strength. In the symmetric gauge, the tunneling matrix elements take
the following values: $V_{1\sss{L}}=|V_{1\sss{L}}|e^{i\phi/4}$,
$V_{1\sss{R}}^*=|V_{1\sss{R}}|e^{i\phi/4}$,
$V_{2\sss{R}}=|V_{2\sss{R}}|e^{i\phi/4}$, and
$V_{2\sss{L}}^*=|V_{2\sss{L}}|e^{i\phi/4}$. The phase shift $\phi$
is associated with the magnetic flux $\Phi$ threading the system by
a relation $\phi=2\pi\Phi/\Phi_{0}$, in which $\Phi_{0}=h/e$ is the
flux quantum.\cite{refKonig}
\par
In such a structure, the electric and heat current can be defined as
a change in the number of electrons and the total energy per unit
time in lead-$L$, respectively. Namely, $J^L_e={ie\over
\hbar}\langle[H,N_L]\rangle$ and $J^L_Q={i\over
\hbar}\langle[H_L,H]\rangle$ with
$N_\alpha=\sum_{k,\sigma}c^\dag_{\alpha k\sigma}c_{\alpha k\sigma}$.
With the help of the nonequilibrium Green function technique, the
electric and heat currents can be expressed as\cite{refMeir,refKim}
\begin{eqnarray}
J^L_e&=&{e\over h}\sum_\sigma \int d\omega
\tau_\sigma(\omega)[f_L(\omega)-f_R(\omega)],\notag\\
J^L_Q&=&{1\over h} \sum_\sigma\int d\omega
(\omega-\mu_L)\tau_\sigma(\omega)[f_L(\omega)-f_R(\omega)].\notag\\
\end{eqnarray}
$f_\alpha(\omega)=[\exp{\omega-\mu_\alpha\over k_BT_\alpha}+1]^{-1}$
is the Fermi distribution function of lead-$\alpha$ when each lead
is in thermal equilibrium at temperature $T_\alpha$.
$\mu_{\alpha}={eV_\alpha}$ is the chemical potential shift due to
the applied source-drain bias voltage $V_\alpha$. The transmission
spectral function $\tau_\sigma(\omega)$ is given by the following
expression\cite{refMeir,Gongpe}
\begin{equation}
\tau_\sigma(\omega)=4\mathrm
{Tr}[{\bf{\Gamma}}^L{\bf{G}}^r_\sigma(\omega){\bf{\Gamma}}^R{\bf{G}}^a_\sigma(\omega)].\label{transmission}
\end{equation}
${\bf{\Gamma}}^L$ is a $2\times 2$ matrix, describing the coupling
strength between the two QDs and lead-$L$. It is defined as
$[{\bf{\Gamma}}^L]_{jn}=\pi
V_{j\sss{L}}V_{\sss{L}n}\rho_\sss{L}(\omega)$
($V_{\sss{L}n}=V_{n\sss{L}}^*$). We will ignore the
$\omega$-dependence of $\Gamma^L_{jn}$ since the electron density of
states in lead-$L$, $\rho_\sss{L}(\omega)$, can be usually viewed as
a constant. Similarly, we can define $[{\bf{\Gamma}}^R]_{jn}$. In
Eq. (\ref{transmission}) the retarded and advanced Green functions
in Fourier space are involved. These Green functions can be solved
by the equation-of-motion method. By a straightforward derivation,
we obtain the retarded Green function which are written in a matrix
form
\begin{eqnarray}
{\bf G}^r_\sigma(\omega)=\left[\begin{array}{cc} g_{1\sigma}(z)^{-1} & i\Gamma_{12}\\
  i\Gamma_{21}& g_{2\sigma}(z)^{-1}
\end{array}\right]^{-1}\ \label{green},
\end{eqnarray}
with $z=\omega+i0^+$ and
$\Gamma_{jn}=[{\bf\Gamma}^L]_{jn}+[{\bf\Gamma}^R]_{jn}$.
$g_{j\sigma}(z)=[z-\varepsilon_{j}+i\Gamma_{jj}]^{-1}$ is the
zero-order Green function of the QD-$j$ unperturbed by another QD.
The advanced Green function can be readily obtained via a relation
${\bf G}^a_\sigma(\omega)=[{\bf G}^r_\sigma(\omega)]^\dag$. In this
work, due to the spin independence of the structure parameters, the
Green function and transmission spectral function are spin
degeneracy with $g_{j\sigma}=g_{j}$ and
$\tau_\sigma(\omega)=\tau(\omega)$.
\par
In the linear response regime, we can expand the electric and heat
currents up to the linear terms of a temperature gradient $\delta
T=T_L-T_R$ to a thermoelectric voltage $\delta V=V_L-V_R$. The
transport coefficients $L_{ij}$ are defined by the relations
\begin{equation}
\left (
\begin{array}{ccc}
J^L_e\\
J^L_Q
\end{array}\right )=
\left (
\begin{array}{ccc}
L_{11}&L_{12}\\
L_{21}&L_{22}
\end{array}\right )
\left (
\begin{array}{ccc}
V_L-V_R\\
T_L-T_R
\end{array}\right ).\label{t11}
\end{equation}
and can be expressed in terms of the transport integral $K_n={1\over
h}\int d\omega (-{\partial f\over \partial
\omega})\omega^n\tau(\omega)$ as $L_{11}=e^2K_0$,
$L_{21}=L_{12}T=-eK_1$, and $L_{22}=K_2/T$. Then the linear response
conductance ${\cal {G}}=\lim_{V\rightarrow0}{dJ_e\over dV}=L_{11}$
is given by the equation
\begin{eqnarray}
{\cal {G}}=e^2K_0.
\end{eqnarray}
The thermopower of a QD system in a two-terminal configuration can
be found in an open circuit by measuring the induced voltage drop
across a QD when the temperature difference between two leads is
applied. The thermopower is defined by the relation
\begin{eqnarray}
S=-{\delta V\over \delta T}\mid_{J=0},
\end{eqnarray}
and can be expressed as
\begin{eqnarray}
S=-{1\over eT}{K_1\over K_0}.
\end{eqnarray}
The electronic contribution to the thermal conductance defined by
$\kappa_e={\Delta J_Q\over \Delta T}$ can be expressed by
\begin{eqnarray}
\kappa_e=K_1eS+{K_2\over T}.
\end{eqnarray}

\begin{figure}
\begin{center}\scalebox{0.42}{\includegraphics{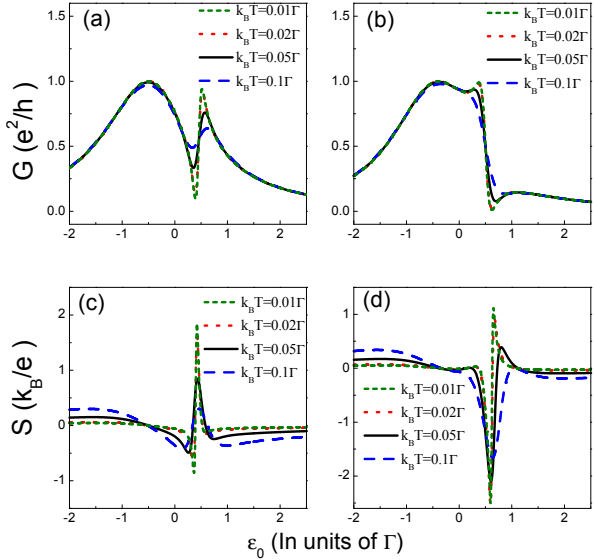}}
\caption{ (a) and (b), the electronic conductances with the increase
of temperature, in the cases of $\phi=0$ and $\phi=\pi$,
respectively. And, the corresponding thermopowers are shown in (c)
and (d). \label{Cond}}
\end{center}
\end{figure}

\section{Numerical results and discussions \label{result2}}
With the formulation developed in the above section, we perform the
numerical calculation to investigate the thermoelectric properties
of the parallel double-QD structure. Prior to the calculation, we
need to introduce a parameter $\Gamma$ as the unit of energy.
\par
According to the discussions in the previous work, this structure is
usually used to research the well-known Fano effect of QD system by
adjusting its arms as nonresonant and resonant channels for electron
transmission. Based on such a result, we here would like to focus on
the thermoelectric effect influenced by the Fano interference in the
parallel double-QD structure. First, we choose
$\Gamma_{22}=10\Gamma_{11}=\Gamma$ to achieve the nonresonant and
resonant transport for the Fano interference. With respect to the QD
levels, we take $\varepsilon_1=\varepsilon_0-{\Gamma\over 2}$ and
$\varepsilon_2=\varepsilon_0+{\Gamma\over 2}$, respectively. In
experiment, $\varepsilon_{0}$ can be changed with respect to the
zero point energy via adjusting the gate voltage. Surely,
$\varepsilon_1$ and $\varepsilon_2$ can be viewed as the levels of
the bonding and antibonding states of the coupled double QDs. In
Fig.\ref{Cond}, we plot the spectra of electronic conductance and
thermopower as functions of $\varepsilon_0$ (i.e., the QD level),
respectively. Fig.\ref{Cond}(a) shows the spectra of the electronic
conductance vs $\varepsilon_0$ in the absence of magnetic field. We
observe that at low temperature, e.g., $k_BT\leq0.02\Gamma$, the
electronic conductance spectrum shows up as a Fano lineshape. With
the increase of temperature, the Fano lineshape in the electronic
conductance spectrum becomes ambiguous. We can understand that such
a result arises from the destructive effect of the increase of
temperature on the Fano interference here. Next, in
Fig.\ref{Cond}(b) we see that the Fano lineshape in the conductance
spectrum is reversed in the case of a magnetic flux through the ring
with $\phi=\pi$. Also, with the increase of temperature the Fano
lineshape becomes unclear. However, from Fig.\ref{Cond}(a)-(b), we
observe that in the two cases that $\phi=0$ and $\phi=\pi$,
temperature plays different roles in modifying the Fano
interference. To be precise, in the zero-magnetic-flux case, the
conductance lineshape is destroyed more seriously by the increase of
temperature. When $k_BT=0.05\Gamma$, the Fano dip in the conductance
spectrum is significantly raised, accompanied by the suppression of
the Fano peak. But, in the case of $\phi=\pi$, the conductance
spectrum is weakly dependent on the increase of temperature, so that
the Fano lineshape can still be seen in the conductance spectrum in
the case that $k_BT=0.05\Gamma$.
\par
In Fig.\ref{Cond}(c)-(d), we investigate the Seebeck effect in the
cases of $\phi=0$ and $\pi$, respectively. From the two figures, we
see that the nonzero Seebeck coefficient only appears in the energy
region where the Fano interference occurs (Hereafter we call such a
region the ``Fano region" for simplicity). Hence, it is evident that
in this structure the Seebeck effect is closely dependent on the
Fano interference. Consequently, at low temperature when the Fano
interference is strong, the magnitude of Seebeck coefficient is
large. And, the increase of temperature weakens the Fano
interference, so that the magnitude of Seebeck coefficient becomes
small. Also due to such a reason, the different Fano lineshapes,
i.e., the different Fano interferences in these two cases cause two
different results about the Seebeck effects. To be specific, in the
zero-magnetic-flux case, the value of Seebeck coefficient is greater
than zero in the Fano region. But when the magnetic flux is applied
with $\phi=\pi$, the sign of the Seebeck coefficient becomes
negative. In addition, it shows that in the case of zero magnetic
field, the magnitude of the Seebeck coefficient is suppressed with
the increase of temperature. Especially when $k_BT=0.1\Gamma$, the
magnitude of $S$ is almost less than $0.5$. However, at the case of
$\phi=\pi$, the temperature dependence of $S$ is weak. Accordingly,
we obtain the result that $S\approx 2$ when $k_BT=0.1\Gamma$.
\par
It should be pointed out that the different thermoelectric
properties in the cases of $\phi=0$ and $\phi=\pi$ originate from
the different Fano interference mechanisms. Next, we try to clarify
the difference between the Fano interferences in such two cases by
employing the concept of Feynman path. With this idea, we rewrite
the electron transmission function as $T(\omega)=4\mathrm
{Tr}[\Gamma^LG^r\Gamma^RG^a]=|\sum\limits_{j, n=1}^2t(j,n)|^2$.
Herein, the electron transmission coefficients are defined as
$t(j,n)=\bar{V}_{\sss{L}j}G^r_{jn}\bar{V}_{n\sss{R}}$ with
$\bar{V}_{j\alpha}=\bar{V}_{\alpha
j}^*=V_{j\alpha}\sqrt{2\pi\rho_\alpha(\omega)}$. Then, we expand the
Green function into an infinite geometric series, e.g.,
$G^r_{11}=\frac{g^{-1}_2}{g^{-1}_1g^{-1}_2+\Gamma_{12}\Gamma_{21}}
=\sum\limits_{j=0}^\infty g_1(-g_1g_2\Gamma_{12}\Gamma_{21})^j$.
Following the expansion of the Green function, the transmission
coefficient $t(1,1)$ can be expressed as a summation of Feynman
paths with different orders, i.e.,
\begin{equation}
t(1,1)=\sum\limits_{j=0}^\infty\bar{V}_{\sss{L}1}
g_1(-g_1g_2\Gamma_{12}\Gamma_{21})^j\bar{V}_{1\sss{R}}=\sum\limits_{j=0}^\infty
t_j(1,1).
\end{equation}
For example, $t_0(1,1)=\bar{V}_{\sss{L}1}g_1\bar{V}_{1\sss{R}}$ is
the zero-order Feynman path from lead-$L$ to lead-$R$ via QD-1. For
the first-order Feynman path, it can be written as
$t_1(1,1)=-\bar{V}_{\sss{L}1}g^2_1g_2\Gamma_{12}\Gamma_{21}\bar{V}_{1\sss{R}}$,
which consists of four terms representing individual Feynman paths.
They are denoted as
\begin{equation}
\left\{
\begin{array}{cccc}
t_{1a}(1,1)&=&-\bar{V}_{\sss{L}1}g_1\widetilde{V}_{1\sss{L}}
\widetilde{V}_{\sss{L}2}g_2\widetilde{V}_{2\sss{L}}
\widetilde{V}_{\sss{L}1}g_1\bar{V}_{1\sss{R}},\\
t_{1b}(1,1)&=&-\bar{V}_{\sss{L}1}g_1\widetilde{V}_{1\sss{L}}
\widetilde{V}_{\sss{L}2}g_2\widetilde{V}_{2\sss{R}}
\widetilde{V}_{\sss{R}1}g_1\bar{V}_{1\sss{R}},\\
t_{1c}(1,1)&=&-\bar{V}_{\sss{L}1}g_1\widetilde{V}_{1\sss{R}}
\widetilde{V}_{\sss{R}2}g_2\widetilde{V}_{2\sss{L}}
\widetilde{V}_{\sss{L}1}g_1\bar{V}_{1\sss{R}},\\
t_{1d}(1,1)&=&-\bar{V}_{\sss{L}1}g_1\widetilde{V}_{1\sss{R}}
\widetilde{V}_{\sss{R}2}g_2\widetilde{V}_{2\sss{R}}
\widetilde{V}_{\sss{R}1}g_1\bar{V}_{1\sss{R}},
\end{array}\right.\label{t11}
\end{equation}
with $\widetilde{V}_{j\alpha}=\widetilde{V}_{\alpha
j}^*=V_{j\alpha}\sqrt{\pi\rho_\alpha(\omega)}$.
\begin{figure}
\begin{center}\scalebox{0.42}{\includegraphics{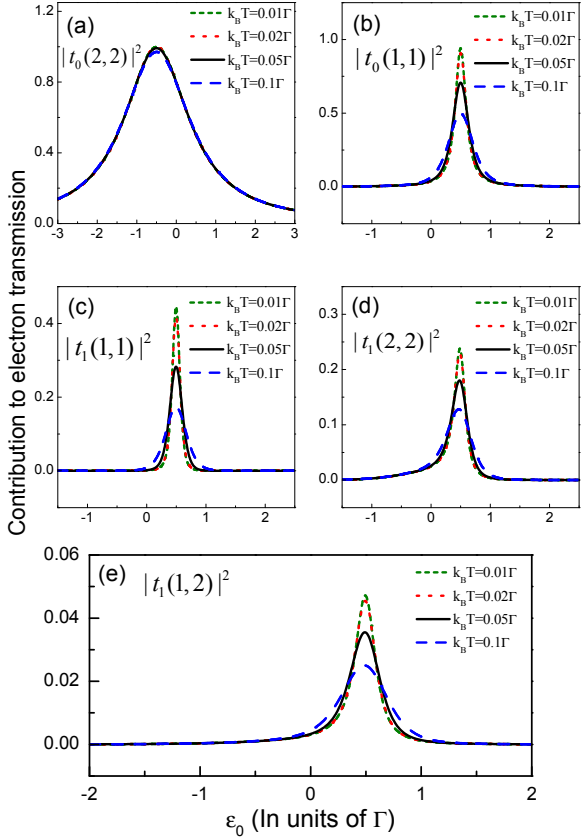}}
\caption{ The contributions of the zero- and first-order Feynman
paths to the electron transmission at the case of zero magnetic
field. \label{Contribut}}
\end{center}
\end{figure}
Similarly, we can expand other transmission coefficients as a
summation of Feynman paths, e.g.,
\begin{eqnarray}
t(1,2)&=&\sum\limits_{j=1}^\infty
i\bar{V}_{\sss{L}1}(-g_1g_2\Gamma_{12})^{j}\Gamma_{21}^{j-1}\bar{V}_{2\sss{R}}
=\sum\limits_{j=1}^\infty t_j(1,2).\notag\\
\end{eqnarray}
The lowest-order Feynman paths of $t(1,2)$ can be written as
\begin{equation}
\left\{
\begin{array}{ccc}
t_{1a}(1,2)&=&-i\bar{V}_{\sss{L}1}g_1\widetilde{V}_{1\sss{L}}
\widetilde{V}_{\sss{L}2}g_2\bar{V}_{2\sss{R}},\\
t_{1b}(1,2)&=&-i\bar{V}_{\sss{L}1}g_1\widetilde{V}_{1\sss{R}}
\widetilde{V}_{\sss{R}2}g_2\bar{V}_{2\sss{R}}.
\end{array}\right.
\end{equation}
By the same approach, the Feynman paths arising from $t(2,2)$ and
$t(2,1)$ can be clarified.
\par
Via the analysis above, we can clearly know that the Fano effect in
this structure originates from the quantum interference among
infinite Feynman paths. But when the magnetic flux is introduced
with $\phi=\pi$, the contribution of the higher-order Feynman paths
will vanish due to the destructive interference among them. Then in
such a case, the two QDs become decoupled from each other with
$\Gamma_{jn}=0$ $(j\neq n)$. And, the interference between
$t_0(1,1)$ and $t_0(2,2)$ leads to the Fano effect. Based on this
viewpoint, we can understand that the different Fano lineshapes in
the electronic conductance spectra arise from the dissimilar Fano
interference mechanisms. Also for such a reason, the thermoelectric
effect in these two cases present different properties. With the
help of the Feynman path method, we next illustrate the sensitive
effect of temperature in the case of $\phi=0$. In
Fig.\ref{Contribut}, we investigate the contributions of the zero-
and first-order Feynman paths to the electronic conductance. For the
zero-order paths, we find that the increase of temperature can not
affect the magnitude of $t_0(2,2)$, whereas the resonant path
$t_0(1,1)$ is destroyed via the increase of temperature. Surely,
such a result brings about the temperature-induced change of Fano
effect in the case of $\phi=\pi$. Next, in
Fig.\ref{Contribut}(c)-(e) one sees that by the increase of
temperature, the contributions of the first-order paths are weakened
seriously. This is because that the temperature increase destroys
the coherent transmission in the paths. According to this result, we
can ascertain that the contributions of the higher-order paths will
decrease seriously with the increase of temperature, since the
complication of high-order paths. Thus, even if the same increase of
temperature, the quantum interference of $\phi=0$ will be further
weakened, compared with the case of $\phi=\pi$. Up to now, we have
understood the sensitive effect of temperature on decrease of
thermoelectric efficiency in the zero-magnetic-flux case.

\begin{figure}
\begin{center}\scalebox{0.42}{\includegraphics{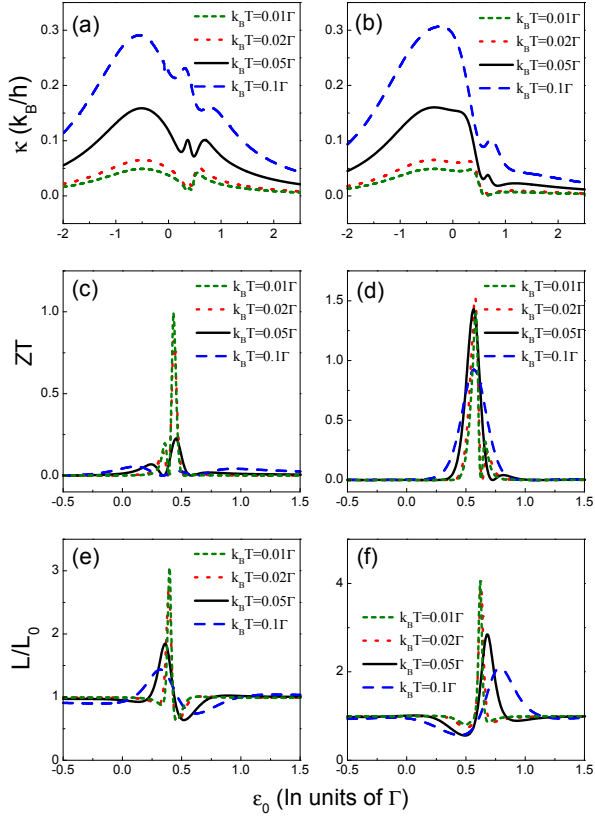}}
\caption{ (a) and (b) The thermal conductances affected by the
temperature increase, in the cases of $\phi=0$ and $\phi=\pi$,
respectively. The corresponding thermal efficiencies are
respectively shown in (c) and (d), whereas the Lorenz numbers are in
(e) and (f). \label{Tcond}}
\end{center}
\end{figure}
\par
In the following, we investigate the thermal conductance $\kappa$,
figure of merit $ZT$, and Lorenz number $L$ in this structure,
respectively. The numerical results are shown in Fig.\ref{Tcond}.
First, in Fig.\ref{Tcond}(a)-(b) we show the thermal conductance
spectra in the cases of $\phi=0$ and $\phi=\pi$, respectively. In
these two figures, we observe that only at low temperature (e.g.,
$k_BT\leq 0.02\Gamma$), the thermal conductance spectra are similar
to those of electronic conductance. But, the temperature dependence
of thermal conductance is more sensitive compared with the
electronic conductance, especially in the case of zero magnetic
flux. Typically, when temperature increases to $k_BT=0.05\Gamma$,
the value of thermal conductance is raised and a sub-peak emerges in
the Fano region of the thermal conductance spectrum. Alternatively,
for the case of $\phi=\pi$, in the Fano region the thermal
conductance increases a little with the increase of temperature to
$k_BT=0.05\Gamma$. Next, Fig.\ref{Tcond}(c)-(d) show the spectra of
figure of merit $ZT$ with the change of QD level. Here we see that
in the zero-magnetic-flux case, the magnitude of $ZT$ is always less
than one. And, with the increase of temperature, the value of $ZT$
decreases to a large degree. For instance, in the case of
$k_BT=0.1\Gamma$, the value of $ZT$ is almost close to zero. But
when $\phi=\pi$, the thermoelectric efficiency is enhanced to a
large degree in the Fano region. Even if $k_BT=0.1\Gamma$, the
magnitude of $ZT$ is close to one. Besides, we see that in such a
case, the figure of merit is nearly independent of the increase of
temperature at low temperature of $k_BT\leq 0.05\Gamma$. Now, we can
clarify the difference of the thermoelectric properties in these two
cases.
\par
It is known that in the bulk materials the thermal and electric
conductances at low temperature are related by the Wiedemann-Franz
law, i.e., $\kappa/{\cal G} T=L_0$, where $L_0= k_B^2\pi^3/3e^2$.
Then, in order to investigate the violation of Wiedemann-Franz law
in such a structure, we reassume $L=\kappa/{\cal G} T$ and evaluate
the value of $L/L_0$ in Fig.\ref{Tcond}(e)-(f). It is seen that in
both cases of $\phi=0$ and $\phi=\pi$, the Lorenz number is deviated
from its classical value in the Fano region. And, at the same
temperature, the value of $L/L_0$ in the case of $\phi=\pi$ is much
larger than that in the case of $\phi=0$. In addition, it is obvious
that with the increment of temperature, the magnitude of $L/L_0$
decreases since the weakness of the Fano interference effect.
\par
We readily find that all the configurations of coupled double QDs
can be mapped into the above model by means of representation
transformation. So, the thermoelectric properties of the above model
can help to further understand those of coupled double-QD systems.
On the other hand, in order to clarify the thermoelectric properties
of coupled double QDs, it is necessary for us to transform its QD
Hamiltonian into the molecular orbital representation. We take the
parallel coupled double QDs as an example to illustrate such an
issue. The single-electron Hamiltonian of it is given by
$h=\underset{k,\alpha\in L,R}{\sum }\varepsilon _{\alpha k}c_{\alpha
k}^\dag c_{\alpha k}+\sum_{j=1}^{2}e _{j}f_{j}^\dag f_{j}+\lambda
f^\dag_1f_2+\underset{\alpha k j}{\sum } w_{j\alpha}f_{j}^\dag
c_{\alpha k }+{\mathrm {h.c.}}$. $f^{\dag}_{j\sigma}$
$(f_{j\sigma})$ is the creation (annihilation) operator of electron
in QD-$j$. $e_j$ denotes the electron level in the corresponding QD.
$\lambda$ is the interdot coupling. $w_{j\alpha}$ denotes the
coupling between the QDs and leads, which is real in the absence of
magnetic flux. As reported by the some works, such a structure
possesses interesting thermoelectric properties.\cite{Trocha} Let us
analyze this structure by mapping the QD Hamiltonian into its
molecular orbital representation. We obtain the relation between QD
level $e_j$ and eigenlevel $\varepsilon_j$, i.e.,
$\varepsilon_{1}=\frac{1}{2}\bigg(e_1+e_2-\sqrt{(e_1-e_2)^2+4\lambda^2}\bigg)$
and
$\varepsilon_{2}=\frac{1}{2}\bigg(e_1+e_2+\sqrt{(e_1-e_2)^2+4\lambda^2}\bigg)$.
Besides, the couplings between the molecular states and
lead-$\alpha$ are expressed as $ \left [
\begin{array}{ccc}
V_{1\alpha}\\
V_{2\alpha}
\end{array}\right ]=
[\bm\eta] \left [
\begin{array}{ccc}
w_{1\alpha}\\
w_{2\alpha}
\end{array}\right ]
$ with
\begin{equation}
[\bm\eta]=\left [
\begin{array}{ccc}
\sqrt{\frac{\lambda^2}{\lambda^2+(\varepsilon_1-e_1)^2}}&\sqrt{\frac{(\varepsilon_1-e_1)^2}{\lambda^2+(\varepsilon_1-e_1)^2}}\\
-\sqrt{\frac{\lambda^2}{\lambda^2+(\varepsilon_2-e_1)^2}}&\sqrt{\frac{(\varepsilon_2-e_1)^2}{\lambda^2+(\varepsilon_2-e_1)^2}}
\end{array}\right ].\notag
\end{equation}
With the above results, the coupling strengths between the molecular
states and the leads can be evaluated, i.e.,
$\Gamma^\alpha_{jj}=\pi|\eta_{j1}w_{1\alpha}+\eta_{j2}w_{2\alpha}|^2\rho_\alpha(\omega)$
and
$\Gamma^\alpha_{jl}=\pi(\eta_{j1}w_{1\alpha}+\eta_{j2}w_{2\alpha})(\eta^*_{n1}w^*_{1\alpha}+\eta^*_{n2}w^*_{2\alpha})\rho_\alpha(\omega)$.
It is certain that the couplings between the molecular states and
the leads determine the quantum transport properties of coupled
double QDs. When changing $e_j$ and $w_{j\alpha}$ to satisfy the
condition of $\Gamma_{jj}\ll \Gamma_{nn}$, we achieve the resonant
and nonresonant channels for electron transmission, so that the Fano
effect will occur. One will then observe the enhancement of
thermoelectric efficiency. Such a result can be attributed to the
destructive quantum interference among infinite Feynman paths.
Moreover, for a typical structure with $e_j=\varepsilon_0$,
$\lambda={\Gamma\over 2}$, and $|w_{j\alpha}|=2|w_{n\alpha}|$, there
will be $\Gamma_{jj}\approx10\Gamma_{nn}$. Then, in the
configuration of left-right symmetry, i.e., $w_{jL}=w_{jR}$, the
quantum interference and the thermoelectric behaviors just
correspond to the case of $\phi=0$ in our model. Alternatively, for
a structure with $w_{jL}=w_{nR}$, there will be $\Gamma_{jn}=0$
($j\neq n$). The thermoelectric properties will be the same as those
in our model of $\phi=\pi$, which is caused by the Fano interference
which occurs between two zero-order Feynman paths. Up to now, by
analyzing the quantum interference of the coupled double-QD
structure, we have clarified its thermoelectric features.
\par
At last, we would like to state that the theory in this work can be
generalized to discuss the thermoelectric effect of the multi-QD
structures. From the previous literature,\cite{Orellana,jpcm,pap} we
know that multi-QD systems possess abundant quantum interference
mechanism. And in such systems, the couplings between some molecular
states and leads act as nonresonant channels while the states
provide resonant channels for electron transmission, so the Fano
effect comes into being. Besides, since so many tunable structure
parameters, in multi-QD systems the Fano effect is more intricate
compared with double-QD structures. Therefore, we are sure that the
thermoelectric properties of these structures are of much interest.
With the help of our analysis, we here readily discuss the
thermoelectric results of multi-QD structure modulated by the Fano
effect. By mapping the QD Hamiltonian into its molecular orbital
representation, we can first clarify the feature of Fano effect. In
the case of $\Gamma_{jn}\neq 0$ ($j\neq n$), the Fano effect arises
from the quantum interference among infinite Feynman paths. And in
such a case, thermoelectric quantities are more sensitive to the
change of temperature, compared with the structure of double QDs.
This is because that in multi-QD structures, the Feynman paths
become more complicated. But when the coupling manners between the
molecular states are $\Gamma_{jn}=0$ ($j\neq n$), it is certain that
the Fano effect is caused by the quantum interference among the
zero-order Feynman paths. Then, similar to the double-QD structure,
the temperature dependence of the thermoelectric quantities is
relatively weak. Therefore, the analysis about the thermoelectric
properties of the double QDs can be generalized to multi-QD case.
Therefore, we believe that our work is helpful for the understanding
about the thermoelectric properties of coupled-QD systems which are
induced by the Fano effect.

\section{summary\label{summary}}
To sum up, in this paper we have discussed the thermoelectric
properties assisted by the Fano effect in a parallel double QD
structure. By adjusting the the couplings between the QDs and leads,
we facilitated the nonresonant and resonant channels for the Fano
interference. It was found that at low temperature, Fano lineshapes
appear in the electronic and thermal conductance spectra. And, the
Fano lineshapes can be reversed by applying a local magnetic flux
with the magnetic phase factor $\phi=\pi$. It showed that the Fano
effect contributes nontrivially to the enhancement of the
thermoelectric efficiency. Furthermore, in the cases of $\phi=0$ and
$\phi=\pi$, the different-property Fano interferences induced the
different thermoelectric effects. Namely, by the presence of
magnetic flux with $\phi=\pi$, the thermoelectric effect is much
more apparent compared with that in the zero-magnetic-flux case. By
employing the concept of Feynman path, the physics reason was
clarified. To be concrete, the Fano effect of $\phi=0$ is caused by
the quantum interference among infinite Feynman paths. Hence, it is
sensitive to the temperature increase, since the increase of
temperature can effectively suppress the coherent transmission in
each path. In contrast, the Fano interference of $\phi=\pi$,
resulting from the interference between two zero-order paths, is
less dependent on the temperature increase. Next, at the end of the
text, we discussed the feasibility of our theory in multi-QD
structures. We hope that the theory of this work is helpful for
understanding the thermoelectric properties contributed by the Fano
effect of coupled-QD systems.

\clearpage

\bigskip

\end{document}